\documentstyle[prd,aps,12pt]{revtex}

\begin{document}

\draft
\title{Model-Independent Extraction of the $N^*(1535)$ Electrostrong
Form Factor from  Eta Electroproduction }
\author{M. Benmerrouche$^{a,b}$,
Nimai C. Mukhopadhyay$^{a}$ and
J.-F. Zhang$^a$ \\
$^a$ Physics Department\\ Rensselaer Polytechnic Institute\\
Troy, NY 12180-3590 \\
$^{b}$ Saskatchewan Accelerator Laboratory \\ University of Saskatchewan\\
Saskatoon, SK S7N 5C6, Canada}

\maketitle

\begin{abstract}

We analyze the existing data on electroproduction of eta 
mesons in the region of W$\approx 1.5$GeV, and extract an
electrostrong form factor for the $N^*(1535)$ electroexcitation
and  decay into the $\eta-N$ channel, which is found to be
relatively insensitive
to the uncertainties of the effective Lagrangian approach.
This  extracted quantity 
is of interest in the QCD description of
relevant baryons.
\end{abstract}



One of the basic questions in baryon physics is how 
an N to N$^*$ electroweak excitation amplitude (N, nucleon, N$^*$,
a nucleon resonance) evolves as a function of the four-momentum
transfer squared, $-Q^2$. The real photon point, $Q^2 =0$,
and the region of relatively low $Q^2$ are clearly the 
domains of non-perturbative QCD. This region is theoretically
difficult to describe, and is currently treated in a variety 
of QCD-inspired models\cite{ref1}. Rigorous calculations
in the lattice QCD framework\cite{ref2} are still in the primitive
stage. At some high enough $Q^2$, the value of which is
under constant debate, the perturbative QCD
scaling rules would set in. Eventually gluon
effects would become visible as scaling violations\cite{ref3}.
There is some crude experimental evidence\cite{ref4}
suggesting the onset of scaling around 4-6 GeV$^2$
region of $Q^2$. The high-Q$^2$ region should also
exhibit the phenomenon of the Bloom-Gilman duality\cite{ref5},
which is a relation between the structure functions of the
resonance and the deep inelastic regions. All these 
theoretical expectations provide a dramatic setting for the
excited baryon  studies at the newer generation ``continuous
wave" (cw) facilities for electrons, such as CEBAF at the
Jefferson Lab, where polarized targets and beams would be 
available for such studies.

This brings us to the subject of this Letter, the process
\begin{equation}
e+p\rightarrow e^{\prime}+p+\eta ,
\end{equation}
in the region of the cm energy W$\approx 1.5 GeV$,
corresponding to the excitation of N$^*$(1535), the so-called
S11 resonance, with $J^{\pi}I={\frac{1}{2}}^{-}\frac{1}{2}$.
A fair bit of data on  this reaction exists\cite{ref6}
from the experiments at the older generation accelerators. 
Some precise real photon studies\cite{ref7} have been
recently done at Mainz.
 While we await   more precise  
experiments at CEBAF\cite{ref8},
 the older data set can already give us valuable insights
in the electrostrong amplitude, characteristic of the excitation and
decay of the $N^*(1535)$ resonance. This is what we intend to do
here.
 
 Using existing data  on (1) and 
an effective Lagrangian approach\cite{ref9}, we shall show
that nearly model-independent inference on the product of the
transverse helicity amplitude and the strong decay 
amplitude  is possible. This, together
with the study of the strong decay property of the N$^*$(1535)
 at hadron facilities like SATURNE\cite{ref10} and
 COSY, would eventually allow us to examine the 
behavior of the transverse helicity amplitude $A_{1/2}$
{\it alone} as a function of $Q^2$. The quantity extracted by
us is of direct interest to the QCD structure of the relevant baryons,
viz., nucleon and N$^*$(1535).

We note at the outset that the reaction (1) is completely dominated
by the N$^*$(1535) resonance (Fig.1). This resonance is best
looked at via the $\eta-N$ channel, as the latter is rather
remarkable in {\it avoiding} a strong coupling to other N$^*$ states, in
contrast to N$\pi$, which exhibits the  property
for strong coupling to many N$^*$'s. Thus, the theoretical interpretation
becomes much simpler in the p$\eta$ decay of the N$^*$(1535),
in contrast to the p$\pi$ decay channel.

The most general effective Lagrangian for the $\gamma NN^{*}$(1535)
vertex is, with $R=N^{*}(1535)$,
\begin{equation}
{\cal L}^{1}_{\gamma NR}=\frac{e}{2(M_{R}+M)}\bar{R}(G^{s}_{1}(k^{2}
)+G_{1}^{v}(k^{2})\tau_{3})\gamma_{5}\sigma_{\mu\nu}NF^{\mu\nu}+
H.c.,
\end{equation}
\begin{equation}
{\cal L}^{2}_{\gamma NR}=\frac{e}{(M_{R}+M)^{2}}\bar{R}(G_{2}^{s}
(k^{2})+G^{v}_{2}(k^{2})\tau_{3})\gamma_{5}\gamma_{\mu}N\partial_{
\nu}F^{\mu\nu}+H.c.,
\end{equation}
taking the pseudoscalar  coupling at the $\eta NN^{*}$(1535)
vertex, where $F^{\mu\nu}$ is the electromagnetic field tensor,
$s$ and $v$ are superscripts indicating isoscalar and isovector
transition form factors, which are unknown, to be determined
from a fit to the existing data\cite{ref6} on the differential 
cross section.  $M_R$ and $M$ are the relevant baryon masses.
The kinematics for the virtual photon
  four momentum $k= (k_{0}$,$\vec{k}$) is the usual one:
$k^{2}\equiv -Q^{2}=(k_{1}-k_{2})^{2}\approx - 4E_{1}E_{2}\sin^{2}
\psi /2$ ,
$\psi$ is the electron scattering angle, $E_{1}$, $E_{2}$,
$\vec{k}_{1}$, $\vec{k}_{2}$ are energies and momenta of the
incident and scattered electrons. The S matrix for the
process (1) is
\begin{equation}
S_{fi}=\frac{e}{(2\pi)^{7}}\delta^{4}(p_{f}+k_{2}+q-p_{i}-
k_{1})\sqrt{\frac{m^{2} M^{2}}{2\omega E_{1}E_{2}E_{i}E_{f}}}
i{\cal M}_{fi}.
\end{equation}
Here $m$ is the meson mass; the hadron four-momenta are, for the incoming and 
outgoing nucleons, $p_{i}=(E_{i}, -\vec{k})$, $p_{f}=(E_{f}, 
-\vec{q})$, and for the $\eta$ meson, $q=(\omega ,\vec{q})$,
in the cm frame of the final nucleon and the meson, defined by
$\vec{q}+\vec{p}_{f}=\vec{k}+\vec{p}_{i}=0$. For the lack of
space, we omit the Born terms for the non-resonant meson
production\cite{ref9}, and give below the expressions for $i
{\cal M}_{fi}$ for the s-channel excitation of the resonance R, using
the Lagrangian in (2) and (3):
\begin{equation}
i{\cal M}_{fi}^{1}=\frac{eg_{\eta}G_{1}^{p}(k^{2})}{(M+M_{R})
}\bar{U}_{f}\frac{\gamma\cdot (p_{i}+k)+M_{R}}{s-M_{R}^{2}}
\gamma_{5}\gamma\cdot k\gamma\cdot\epsilon U_{i},
\end{equation}
\begin{equation}
i{\cal M}_{fi}^{2}=\frac{eg_{\eta}G_{2}^{p}(k^{2})k^{2}}
{(M+M_{R})^{
2}}\bar{U}_{f}\frac{\gamma\cdot (p_{i}+k)+M_{R}}{s-M_{R}^{2}}
\gamma_{5}\gamma\cdot\epsilon U_{i},
\end{equation}
with $g_{\eta}$, the $\eta NR$  coupling, $U_{i}$ and 
$U_{f}$, the spinors for incoming and outgoing $N$, $s=W^{2}=(E_{i}+k_{0})^{2}$.
Note that the second term vanishes for the real photon. For the u-channel,
the amplitude can be constructed by crossing symmetry.

The canonical procedure for calculating the differential 
cross section for the process and polarization observables,
 is to write ${\cal M}_{fi}$ in terms of the
CGLN-type\cite{ref11} amplitude $\cal F$:
${\cal M}_{fi}=(4\pi W/M)\chi_{f}^{\dagger}{\cal F}\chi_{i}$,
where the $\chi_{i}$ and $\chi_{f}$ are the nucleon Pauli
spinors, taking into account the transitions $\gamma N
\rightarrow N^{*}\rightarrow\eta N$, where $\gamma$ is
the virtual photon. The amplitude $\cal F$ is given by
${\cal F}  = i\vec{\sigma}\cdot\vec{b}{\cal F}_{1}+\vec{\sigma}
\cdot\hat{q}\vec{\sigma}\cdot (\hat{k}\times\vec{b})
{\cal F}_{2}+i\vec{\sigma}\cdot\hat{k}\hat{q}\cdot\vec{b}
{\cal F}_{3}
 +i\vec{\sigma}\cdot\hat{q}\hat{q}\cdot\vec{b}{\cal F}_{4}-
i\vec{\sigma}\cdot\hat{q}b_{0}{\cal F}_{5}-i\vec{\sigma}
\cdot\hat{k}b_{0}{\cal F}_{6}$,
with 
$b_{\mu}=\epsilon_{\mu}-(\vec{\epsilon}\cdot\hat{k}/|
\vec{k}|)k_{\mu}$.
The ${\cal F}_{i}$'s can be converted into helicity amplitudes
$H_{i}$ ($i=1$, ... 6),
in terms of which the differential cross section can be 
written appropriately:
\begin{equation}
\frac{d\sigma}{d\Omega}=\frac{d\sigma_{T}}{d\Omega}+\epsilon
\frac{d\sigma_{s}}{d\Omega}+\epsilon \cos 2\phi\frac{d
\sigma_{p}}{d\Omega}+\sqrt{2\epsilon (1+\epsilon )}\frac{d
\sigma_{I}}{d\Omega}\cos\phi,
\end{equation}
wherein various structure functions of the right-hand side
can be rewritten in terms of the bilinears of the helicity
amplitudes. In  (7), $\phi$ is the azimuth and 
$\epsilon$ is the virtual photon polarization\cite{ref12}. 
We can express the helicity amplitudes in terms of the multipole
amplitudes as well. In the $N\rightarrow N^{*}$(1535)
case, we have to deal with two helicity amplitudes $A_{1/2}$ and 
$S_{1/2}$, which can be given in terms of $G_{1}^{p}$ and 
$G_{2}^{p}$ in (5), and (6).

Our procedure to fit the existing differential cross-section
data\cite{ref6} is the following. We fix the Born terms for
nucleon and vector meson exchanges as in the real photon
case\cite{ref9}, except for the form factors. The 
nucleon form factors have the usual dipole form, while the
$\rho\eta\gamma$ and $\omega\eta\gamma$ electromagnetic
form factors are parametrized in terms of the prescription
of the vector dominance\cite{ref13}. Thus,
$G_{V\gamma\eta}(k^{2})=(1-k^{2}/m^{2}_{V})^{-1}$,
where $m_{V}\approx\frac{1}{2}(m_{\rho}+m_{\omega})$, 
the average vector meson mass. It is a reasonable 
approximation to neglect relatively small contributions
from  nucleonic resonances, such as D13(1520), to the
angular distributions at the crude level of precision of
the old data. However, high precision of data expected
in new facilities and polarization observables would require
their inclusion. With the existing data base on electroproduction
of etas, it is not possible to extract any meaningful
information on other resonances. Given the relative importance
of the nucleon Born terms, vector meson exchanges and the
excitation of N$^*$(1535) in the ascending order, we use this
model to determine the $A_{1/2}(Q^{2})$, given some Ans\"{a}tze for the
small scalar (longitudinal)\footnote{There are different 
conventions\cite{ref14,ref15,ref16,ref17}
involved in the definition of scalar (or, equivalently, longituginal)
helicity amplitude.} 
 amplitude $S_{1/2}(Q^{2})$. Since the current
experimental data are not accurate enough to pin down the
longitudinal strength of $S_{11}\rightarrow \gamma +N$
transition, we have chosen three scenarios for the value of the
ratio $R_{LT}=S_{1/2}/A_{1/2}$:      
(a)  $R_{LT}=0$; (b) fix $R_{LT}$ by the quark
shell model\cite{ref15}; (c) use the value of $R_{LT}$
from the works cited as Refs. [14, 15, 16] in Stanley and Weber\cite{ref16}.
 This gives us a measure of the
uncertainty in extracting the tranverse helicity amplitude,
given that for the longitudinal amplitude.

In Fig.1, we show the angular distributions  measured\cite{ref6} in the 
reaction (1) and our best fits in the effective Lagrangian approach. Notice
the dominance of the N$^*$(1535) excitations: As
we turn off the N$^*$(1535) contribution, the differential
cross section collapses completely.  Thus, it makes sense to extract
the electrostrong property of the $N^*(1535)$ resonance from the
process (1).

In Table I, we give the value of the parameter 
$\xi_{T}=\sqrt{\chi^{\prime}\Gamma_{\eta}}A_{1/2}/\Gamma_{T}$
where $\chi^{\prime}$ is a kinematic parameter\cite{ref9},
$\sqrt{\Gamma_{\eta}}/\Gamma_{T}$ is the $N^* \rightarrow
p\eta$ decay amplitude. This parameter\footnote{In Table 1, we
are  including  only results from old data base for $Q^2 =0$.
New photoproduction data from Mainz\cite{ref7} 
yield $\xi_{T}=2.20\pm 0.15$ in units of $10^{-1}GeV^{-1}$.
The definition of $\xi_{T}$ contains the kinematic parameter
$\chi^{\prime}$ so that it is consistent with the definitions of
the Particle Data Group\cite{ref18}.}
 is our extracted
electrostrong form factor for the $N^*(1535)$ resonance,
of interest to the QCD description of baryons.
We note the relative insensitivity of this quantity 
to a variation of parameters of the model inputs, such as the
resonance parameters, value of $g_{\eta pp}$, vector meson
form factor and so on.
 This is the {\it central result} of our Letter.

In Fig. 2, we plot $\xi_{T}$ for 
different inputs of the $S_{1/2}$ to $A_{1/2}$ ratio.
This shows relative insensitivity of the extracted parameter
$\xi_{T}$ to the current experimental and theoretical
uncertainties in the extraction of the longitudinal to 
transverse amplitude ratio. We also include the prediction
of a light front approach 
from Stanley and Weber\cite{ref16}. The present nonrelativistic
versions of the
quark  model\cite{ref15}, predictions of which
are represented by the dot-dashed lines, and the prediction
from Stanley and Weber\cite{ref16}   are unable to reproduce the
variation of this extracted parameter as a function of
$Q^2$.

In summary, measured angular distributions of the $\eta$
electroproduction process allow us, in the effective 
Lagrangian approach, to extract the form factor characteristic of the
$\gamma pN^{*}$  and $\eta pN^{*}$ vertices,
which is essentially model independent.  The current
versions of quark model, though quite successful in
phenomenological terms,  are unable to explain the
$Q^2$ dependence of this extracted electrostrong
form factor.  
Thus, we urgently need rigorous non-perturbative calculations
using QCD on the lattice.

This work has been mostly done at RPI, where the authors have been
supported by the U.S. Department of Energy. The work at
Saskatoon has been supported by the Natural Sciences and Engineering Research
Council of  Canada. One of us (N.C.M.) thanks Dr. B. Saghai for his hospitality
at CEN, Saclay, and for much enjoyable discussions.



\newpage

Fig. 1: {Angular distributions for eta mesons 
  and our best fits (solid line) in the effective Lagrangian
approach. The dashed line is without N$^*$(1535). The data are from Ref.6.} 

Fig. 2: {$\xi_{T}$ vs. $Q^2$ for different prescriptions of $S_{1/2}$
to $A_{1/2}$ ratio: (a) set $S_{1/2}=0$ ( circles connected by a solid
line); (b) fix $S_{1/2}/A_{1/2}$
by the quark shell model\protect{\cite{ref15}}
 ( squares connected by a dashed line); (c) 
use the value of $S_{1/2}$ from refs. [14, 15, 16] of ref. [16] 
(diamonds connected by a dotted line).
The non-relativistic quark model prediction of Ref. [15] is the dot-dashed
line. The prediction from a light front approach of Stanley and 
Weber\protect{\cite{ref16}} is also shown (long-dashed line)
 with their parameter
$\alpha =0.2 GeV^{2}$.}

\newpage

\begin{table}
\caption{The fitted results of $A_{1/2}$ and $\xi_{T}$ for
different models ($S_{1/2}=0$ here).
 Model 1 and 2 are with different mass positions
 widths and decay ratios ($W=1535$ MeV, 1549 MeV, $\Gamma=150$ MeV, 202 MeV,
$\Gamma_{\eta}/\Gamma$=0.5 and 0.55 respectively). Model 3 is
the result of doubling $\eta$-nucleon coupling constant.
Model 4 is the result of change of the cut-off of form factor
at vector meson nucleon vertex from 1.2 $GeV^2$ to 2.0 $GeV^2$.
For each entry the first line is $A_{1/2}$ in unit of $10^{-3}$
$GeV^{-1/2}$ and the second line is $\xi_{T}$ in unit of $ 10^{-1}$
$GeV^{-1}$.}
 
\label{table1}
\begin{center}
\begin{tabular}{ccccc}
$Q^2$ ($GeV^2$) & Model 1&Model 2&Model 3&Model 4 \\
     \hline
0.0 & $88.83\pm 7.03$ & $97.27\pm 5.62$&$87.07\pm 5.44$&$90.18\pm 5.58$\\
&$2.04\pm 0.16$ &$1.90\pm 0.11 $&$2.00\pm 0.12$&$2.07\pm 0.13$ \\ \hline
0.2&$88.93\pm 5.94$&$97.28\pm 6.57$ & $89.00\pm 5.97$ &
     $86.95\pm 6.07$\\
   &$2.04\pm 0.14$&$1.90\pm 0.13$ &$2.04\pm 0.14$ &
    $1.99\pm 0.14$\\ \hline 
 0.28&$91.56\pm 5.85$&$99.99\pm 6.48$ &$91.54\pm 5.88$ & $89.70\pm
      5.98$ \\
     & $2.10\pm 0.13$ &$1.95\pm 0.13$ & $2.10\pm 0.13$&
      $2.06\pm 0.14$ \\ \hline 
0.4&$91.08\pm 5.91$& $99.27\pm 6.54$& $90.78\pm 5.95$& $89.16\pm
     6.04$\\
   & $2.09\pm 0.14$& $ 1.94\pm 0.13$& $2.08\pm 0.14$&
     $2.04\pm 0.14$\\ \hline 
0.6&$90.95\pm 8.50$& $92.79\pm 8.94$& $88.80\pm 8.59$&$91.95\pm
     8.86$\\
   &$2.08\pm 0.19$& $1.81\pm 0.17$& $2.04\pm 0.20$& $2.11\pm
    0.20$\\ \hline 
1.0&$82.83\pm 7.12$& $89.93\pm 7.87$& $82.67\pm 7.16$& $81.07\pm
    7.30$\\
   &$1.90\pm 0.16$& $1.76\pm 0.15$& $1.89\pm 0.16$&$1.86\pm
   0.17$ \\ \hline 
2.0&$59.75\pm 7.10$& $64.65\pm 7.83$&$59.59\pm 7.14$& $58.25\pm 7.31$\\
   &$1.37\pm 0.16$& $1.26\pm 0.15$& $1.34\pm 0.16$&
    $1.34\pm 0.17$ \\ \hline 
3.0&$52.45\pm 5.32$& $57.04\pm 5.86$& $52.40\pm 5.33$&$51.88\pm 5.39$\\
   &$1.20\pm 0.12$& $ 1.11\pm 0.11$&$1.20\pm 0.12$& $1.19\pm 0.12$\\
\end{tabular}
\end{center}
\end{table}

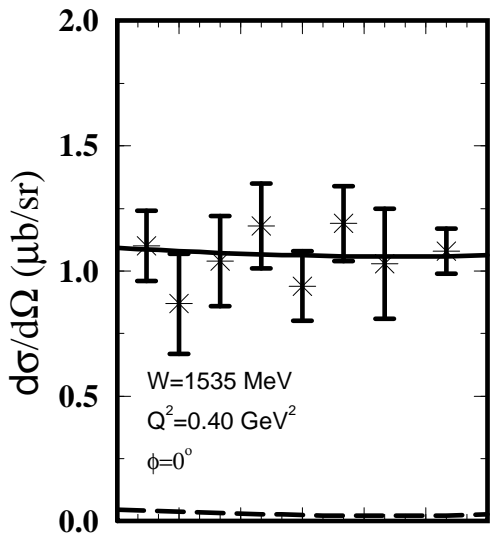
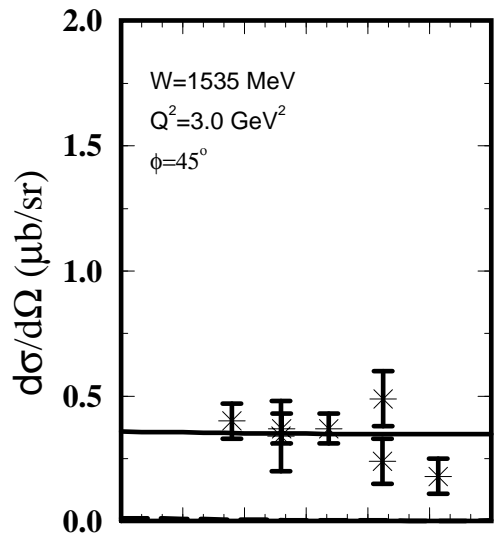
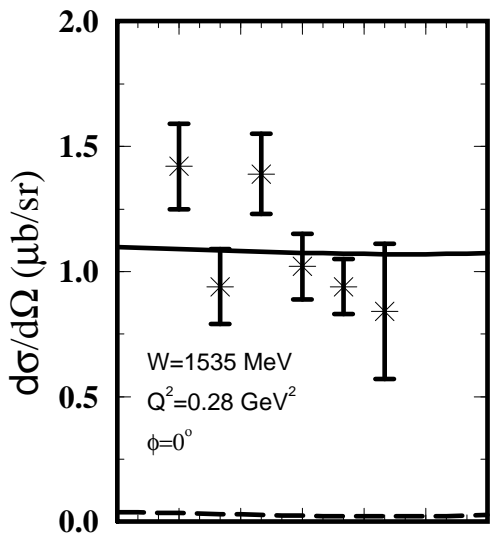
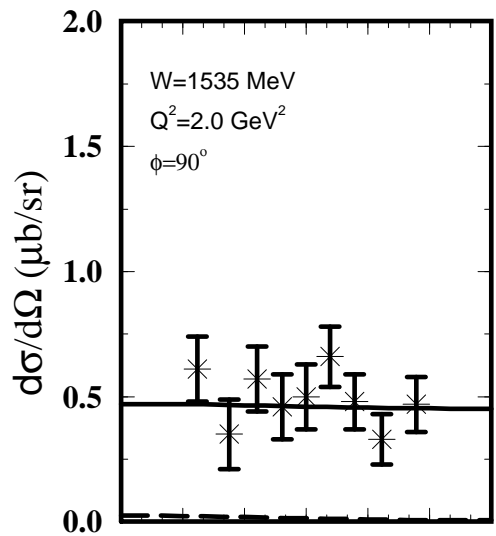
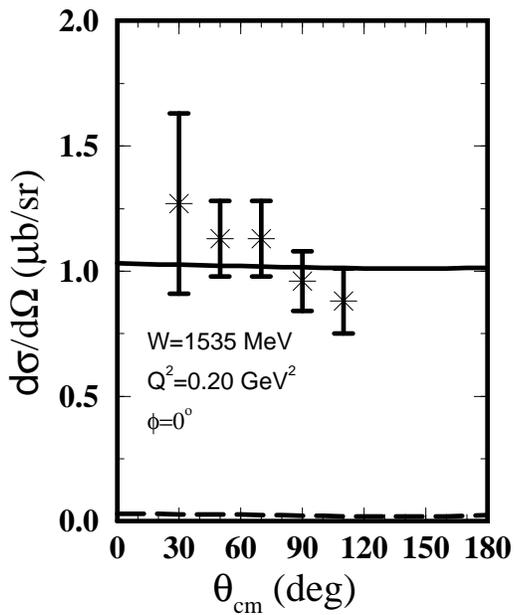
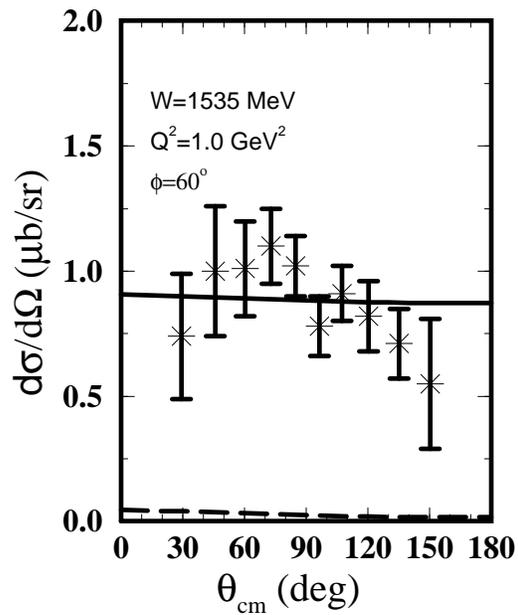
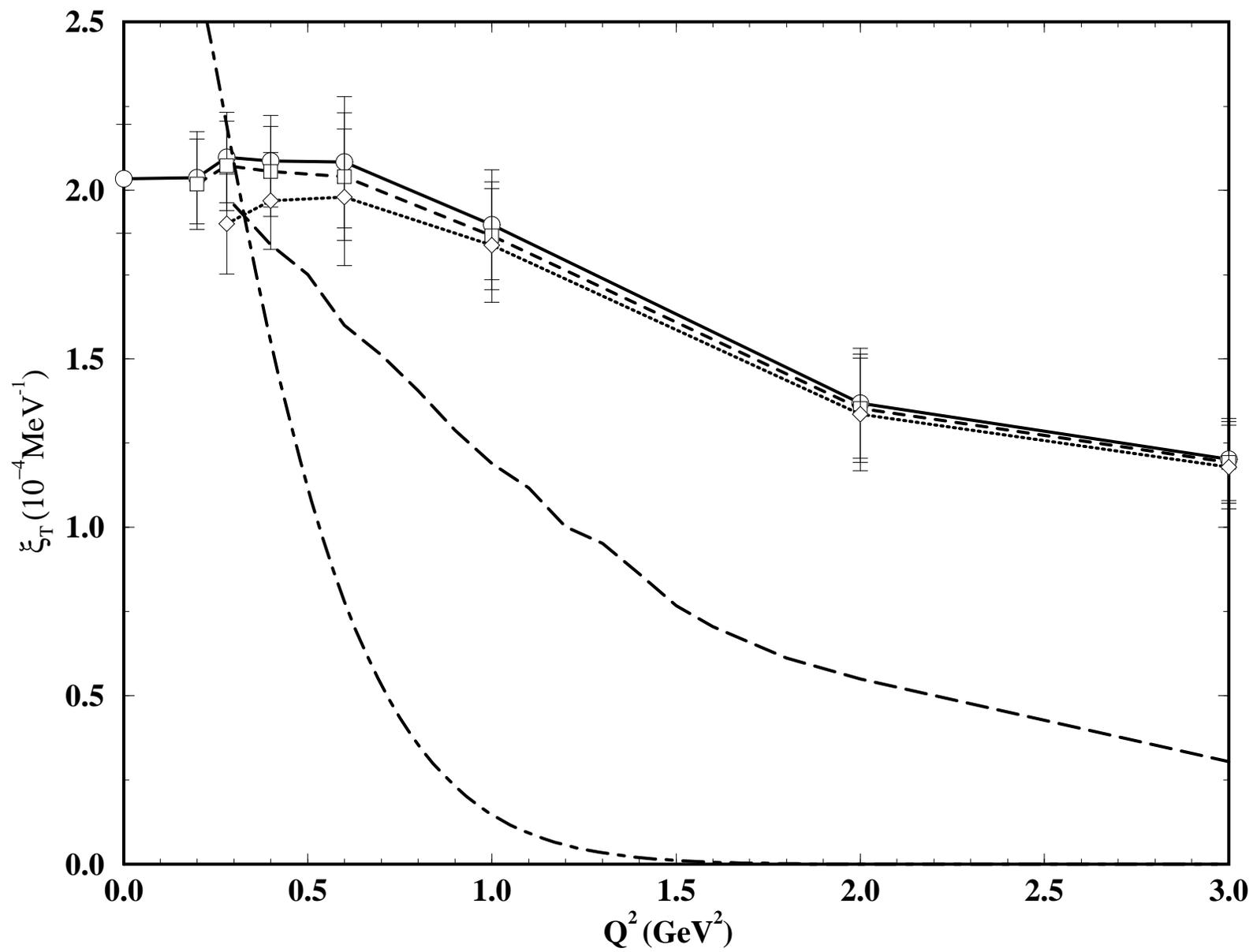
\end{document}